\documentstyle[12pt]{article}
\begin{document}
\pagestyle{empty}
\hfill{ANL-HEP-PR-94-25}
\vskip .5cm
\hfill{ITP-SB-94-33}
\vskip 1cm
\centerline{\Large \bf Non-leading Logarithms in Principal Value
Resummation}
\vskip .5cm
\begin{center}
\begin{tabular}{cc}
Lyndon Alvero & Harry Contopanagos \\
Institute for Theoretical Physics & High Energy Physics Division \\
SUNY Stony Brook & Argonne National Laboratory \\
Stony Brook, NY 11794 & Argonne, IL 60439 \\
\end{tabular}
\end{center}
\vskip 1cm
\centerline{July 1994}
\vskip .5cm
\centerline{\bf Abstract}

We apply the method of principal value resummation of large
momentum-dependent radiative corrections to the
calculation of the Drell Yan cross section.  We sum all next-to-leading
logarithms and provide numerical results for the resummed exponent and the
corresponding hard scattering function.

\newpage
\section{Introduction}
\pagestyle{plain}

All-order resummation techniques have been in existence for
some time now \cite{ref:six,ref:seven}, mainly in an attempt to control
perturbative QCD
corrections to specific hadron-hadron scattering cross sections
which are numerically large at a fixed order of
perturbation theory \cite{ref:eight,ref:nine}.
Such resummation formulas \cite{ref:one} \,
typically organize the large perturbative corrections in an
exponential containing integrals of the QCD running coupling
$\alpha_s(\mu^2)$ over various momentum scales and suffer from
divergences associated with the Landau pole of $\alpha_s(\mu^2)$.

In \cite{ref:two} it was shown, in the context of the inclusive dilepton
cross section, how to define a consistent resummation formula
by a principal value prescription for the infrared (IR) coupling
constant singularities of the exponent.  This defines unambiguously
the resummed perturbative series, as the exact exponent may
be approximated by a finite series in the fixed coupling constant,
thus avoiding an explicit and numerically important presence
of higher-twist contributions through an arbitrary IR cut-off
\cite{ref:three}.   The principal value prescription then defines
a {\em best} perturbative resummation in the following sense:
A perturbative exponent can be constructed as an {\it asymptotic}
approximation to the principal value exponent and its degree
(in $\alpha_s(Q^2)$) is precisely determinable numerically  as a function of
$Q^2$ as well. Thus, the often quoted ``asymptotic nature of the QCD
perturbative series" is given a precise meaning.

In \cite{ref:two}  it was also shown that all
large momentum-dependent perturbative
corrections were contained in leading and next-to-leading
exponents ($E_L$ and $E_{NL}$ respectively). $E_L$ was studied
in that work, both analytically and numerically. $E_{NL}$ is, however,
also important for a precise perturbative description of the process
and a full study of this exponent is necessary before the non-perturbative
parton flux is convoluted with the resummed hard part in order
to yield the theoretical prediction for the cross section.

The purpose of this work is to address these issues and
perform a thorough analytical and numerical study
of the exponentiated hard part, including all large momentum-dependent
perturbative corrections. In a subsequent paper,  we will use
this hard part to calculate the dilepton cross sections at both fixed-target
and collider energies and compare these with experimental results.

This paper is organized as follows:
in section 2 we briefly summarize the new resummation formula
and the expressions for $E_L$, both as an exact function of $Q^2$
and as an asymptotic perturbative series, i.e., as a polynomial
in $\alpha_s(Q^2)$. This makes contact with earlier work \cite{ref:two} and
defines notation and $E_{NL}$ as well.
In section 3 we calculate $E_{NL}$, in the spirit of \cite{ref:two},
both exactly  and as an asymptotic perturbative series.
The results are presented in both moment  and momentum space,
using a closed formula for the inverse Mellin transform, which
has been derived elsewhere \cite{ref:four}.
In section 4 we study the higher-twist inherent in the exact
exponent of the resummation formula. We establish, in moment space,
the asymptotic properties of the exponent in the non-perturbative
region. These properties make evident the smooth behavior of the
cross section for these extremely
large moments.
In section 5 we present numerical results for $E_L$ and $E_{NL}$
in the perturbative and non-perturbative region, along with
the corresponding perturbative approximations.  We also present the
hard part in momentum space, for a variety of values of $Q^2$, including
all large perturbative corrections.
Finally in section 6 we summarize our findings and anticipate
future, phenomenologically oriented work.
 Some technical details are given in an appendix.

\section{Principal-Value Resummation}

In this section we briefly summarize the principal-value
resummation formula as applied to the dilepton production hard scattering
function. Let us define kinematic variables for this
reaction, denoted by
\begin{equation}
h_1(p_1)+h_2(p_2)\to l{\bar l}(Q^\mu)+X\ .
\label{drellyan}
\end{equation}
We denote $s=(p_1+p_2)^2,\ \tau=Q^2/s,\ z=\tau/x_ax_b$ where
$x_a,\ x_b$ are the momentum fractions of the partons participating
in the hard scattering. We shall concentrate on the diagonal-flavor
quark-antiquark hard parts, which alone are singular in the
soft-gluon limit $z\to 1$, at a fixed order of perturbation theory.

For quark-antiquark scattering, we obtain the principal-value resummation
formula in moment space \cite{ref:one,ref:two,ref:four}
\begin{equation}
{\tilde \omega}_{q{\bar q}}(n,\alpha)=A(\alpha){\tilde I}(n,\alpha),
\label{quarkhard}
\end{equation}
where the function $A(\alpha)$ will be fully analyzed in a future
paper, and where $\alpha
\equiv \alpha_s(Q^2)/\pi$.
All $n$-dependence in $\omega_{q {\bar q}}$ is contained in the exponential
\begin{equation}
{\tilde I}(n,\alpha)={\rm exp}[E(n,\alpha)_L+E(n,\alpha)_{NL}]\ .
\label{exponent}
\end{equation}
In eq.~(\ref{exponent}), $E(n,\alpha)_L$ includes the one-loop
running coupling effects, and $E(n,\alpha)_{NL}$ the two-loop effects.
In \cite{ref:two,ref:four} it was shown that, if we define the range of
validity of perturbation theory \footnote{A precise definition
of this range will be discussed below, and in section 4.} by
$\alpha_s(Q^2)\ln n<1$, this formula includes
{\it all} large perturbative corrections. In particular, $E_L$
resums all corrections containing more powers of $\ln n$
than of $\alpha$, while $E_{NL}$ completes the resummation
of corrections having equal powers of these two quantities.
As usual, conclusions in momentum space may be drawn using
the correspondence \cite{ref:four,ref:ten} $n\leftrightarrow 1/(1-z)$.
For $E_L$ in the Deep-Inelastic Scattering (DIS) scheme,
we can write the expressions \cite{ref:two}
\begin{equation}
E(n,\alpha)_L=\alpha(g_1^{(1)}I_1-g_2^{(1)}I_2),
\label{elintegrals}
\end{equation}
with
\begin{equation}
I_1\equiv I_1(n,t)=2I(n,t/2)-I(n,t),
\label{ione}
\end{equation}
where
\begin{equation}
I(n,t)=t\int_Pd\zeta\biggl({\zeta^{n-1}-1\over 1-\zeta}\biggr)
\ln(1+(1/t)\ln(1-\zeta)),
\label{ienti}
\end{equation}
and
\begin{equation}
I_2\equiv I_2(n,t)=\int_Pd\zeta\biggl({\zeta^{n-1}-1\over 1-\zeta}\biggr)
{1\over 1+(1/t)\ln(1-\zeta)},
\label{itwo}
\end{equation}
and where we define
\begin{equation}
t\equiv {1\over \alpha b_2}=\ln\biggl({Q^2\over \Lambda^2}\biggr).
\label{tdef}
\end{equation}
Here $g_1^{(1)},\ g_2^{(1)}$ are numerical coefficients \cite{ref:two},
while $b_2$ is the first coefficient of the QCD beta-function. The
explicit expressions for these coefficients are given in section 5.
$E_{NL}$ is given by a similar, albeit more complicated, expression,
which we will also give below.

The principal value prescription makes the integrals in
eqs.~(\ref{ienti})-(\ref{itwo}) well-defined at \\ \mbox
{$-{1 \over t}\ln (1-\zeta)=1$}.
$P$ stands for an averaged mirror-symmetric
contour connecting 0 and 1 above and below the real axis. It
will be useful to consider two such special contours for
analytical and/or numerical results. The definition of $P$, along
with these  two specific choices, appear in fig. 1.

In \cite{ref:two}, a series expression for the integrals
in eqs.~(\ref{elintegrals})-
(\ref{itwo})\ was found, using the principal value contour proper,
${\cal P}$:
\begin{eqnarray}
E(n,\alpha)_L&=&-\alpha g_1^{(1)}\sum_{\scriptstyle
{m=1}}^\infty{(1-n)_m\over m!m^2}
2{\cal E}(mt/2) \nonumber \\
& &+\alpha g_1^{(1)}\sum_{m=1}^\infty{(1-n)_m\over m!m^2}
{\cal E}(mt) \nonumber \\
& &-\alpha g_2^{(1)}\sum_{m=1}^\infty{(1-n)_m\over m!m}{\cal E}(mt),
\label{elsums}
\end{eqnarray}
where $(a)_m\equiv \Gamma(a+m)/\Gamma(a)$, $\Gamma$ being the Euler
Gamma function, with
\begin{equation}
{\cal E}(x)=x{\rm e}^{-x}Ei(x),
\label{elcalE}
\end{equation}
and where the exponential integral is defined by the principal value
\begin{equation}
Ei(x)\equiv {\cal P}\int_{-\infty}^x dy{\rm e}^y/y.
\label{ieiofx}
\end{equation}

A perturbative version of eq.~(\ref{elsums})\  may be obtained by performing
an asymptotic expansion of the special functions ${\cal E}(mt),{\cal E}(mt/2)$
for $mt, mt/2 \gg 1$ in terms of $\ln^j n/t^\rho$ \cite{ref:two}
and in a range where
the $\ln n$ behavior is suppressed for all terms in the sums\cite{ref:four}.
For example, the second sum in (\ref{elsums}) can be written as
\begin{equation}
\alpha g_1^{(1)}\sum_{m=1}^\infty{(1-n)_m\over m!m^2}{\cal E}(mt)=
\alpha g_1^{(1)}\sum_{\rho=0}^N{\rho! \over t^\rho}
\sum_{m=1}^\infty{(1-n)_m\over m!m^{\rho+2}},
\label{sumex}
\end{equation}
where use had been made of eq. (3.2) of \cite{ref:two}:
\begin{equation}
{\cal E}(mt) \simeq \sum_{\rho=0}^N{\rho! \over (mt)^\rho}.
\label{expexpan}
\end{equation}
Applying eq. (3.8) of \cite{ref:two},
\begin{equation}
\sum_{m=1}^\infty{(1-n)_m \over m!m^{\rho+1}}=(-1)^\rho\sum_{j=0}^{\rho+1}
c_{\rho+1-j}{(-1)^j \over j!} \ln^j n \ ,
\label{lognseries}
\end{equation}
to eq.~(\ref{sumex}), one then obtains
\begin{equation}
\alpha g_1^{(1)}\sum_{m=1}^\infty{(1-n)_m\over m!m^2}{\cal E}(mt)=
\alpha g_1^{(1)}\sum_{\rho=0}^N{\rho! \over t^\rho}(-1)^{\rho+1}
\sum_{j=0}^{\rho+2}c_{\rho+2-j}{(-1)^j \over j!}\ln^j n.
\label{sum2exp}
\end{equation}
 From eq.~(\ref{sum2exp}), we see that for $n\gg 1$, the suppression
of $\ln n$ by $1/t$ can be quantified for a given power $1/t^\rho$, by
constraining the corresponding monomial $\ln^jn,\ j=\rho$.
This leads to $\alpha b_2\ln n<1$ or, equivalently, $n{\rm e}^{-t}<1$,
as the range of validity of perturbation
theory. We will further discuss this argument from a different
point of view, when we investigate the non-perturbative regime in section 4.
Of course, in eq.~(\ref{elsums})\ the sums have two different scales,
the smaller one being $t/2$.
Thus, if we want to define the perturbative regime for all three sums,
we should restrict ourselves to $n$ for which $2\alpha b_2\ln n<1$.
In summary, the perturbative range in $n$ is
\begin{equation}
1<\ln n<{t \over 2}\equiv {1\over 2\alpha b_2}\Leftrightarrow
1\ll n<{Q\over \Lambda}.
\label{perturange}
\end{equation}

For $n$ in this range, a large-$t$ asymptotic approximation
leads to the perturbative formula
\begin{equation}
E(n,\alpha)_L\simeq E(n,\alpha,N)_L=\sum_{i=1}^2\sum_{\rho=1}^{N_i^L+1}
\alpha^\rho\sum_{j=0}^{\rho+1}s_{j,\rho}^L(i)\ln^jn
+\sum_{\rho=1}^{N_3^L+1}\alpha^\rho\sum_{j=0}^\rho s_{j,\rho}^L(3)\ln^j n,
\label{elpert}
\end{equation}
where the $N_i^L$ determine the best asymptotic approximation, and where
the coefficients are given by
\begin{eqnarray}
s_{j,\rho}^L(1)&=&-g_1^{(1)}b_2^{\rho-1}(-1)^{\rho+j}{(\rho-1)!\over
j!}2^\rho c_{\rho+1-j}, \nonumber \\
s_{j,\rho}^L(2)&=&g_1^{(1)}b_2^{\rho-1}(-1)^{\rho+j}{(\rho-1)!\over j!}
c_{\rho+1-j}, \nonumber \\
s_{j,\rho}^L(3)&=&g_2^{(1)}b_2^{\rho-1}(-1)^{\rho+j}{(\rho-1)!\over j!}
c_{\rho-j}.
\label{elcoeff}
\end{eqnarray}
The coefficients $c_k$ are given by $\Gamma(1+z)=\sum_{k=0}^\infty c_kz^k$.
Notice that $E_L$ in eq.~(\ref{elpert})\ becomes a
polynomial in $\ln n$ whose precise powers
are determined by the above large-$t$ asymptotic approximation (perturbative
approximation). Hence, the optimum number of resummed perturbative corrections,
$N\equiv\{N_i^L\}$,
is also determined in principal-value resummation. Numerical results for
those numbers were given in \cite{ref:two} and will be incorporated
and enlarged upon in sec. 5 below.
\section{Evaluation of the Next-to-Leading exponent in the perturbative regime}

Following \cite{ref:two}, we may write the non-leading exponent $E_{NL}$
from eq.~(\ref{exponent}) as an integral over the function
\begin{equation}
g_i^{NL}(\alpha[\lambda Q^2])=-g_i^{(1)}\alpha^2(b_3/b_2)
{\ln(1+\alpha b_2\ln\lambda)\over (1+\alpha b_2\ln\lambda)^2}
+g_i^{(2)}\alpha^2{1\over (1+\alpha b_2\ln\lambda)^2}\ ,\  \ i=1,2 .
\label{ginl}
\end{equation}
In these terms, $E_{NL}$ is
\begin{equation}
E(n,\alpha)_{NL}=-\int_P d\zeta\biggl({\zeta^{n-1}-1\over 1-\zeta}\biggr)
\biggl\{\int_0^\zeta{dy\over 1-y}g_1^{NL}(\alpha[(1-\zeta)(1-y)Q^2])
+g_2^{NL}(\alpha[(1-\zeta)Q^2])\biggr\}\ .\label{enldef}
\end{equation}
 From eqs.~(\ref{ginl}), (\ref{enldef})\ we may write the
next-to-leading exponent as
\begin{equation}
E(n,\alpha)_{NL}=
\alpha(g_1^{(1)}J_1-g_2^{(1)}J_2)+\alpha^2(g_1^{(2)}K_1-g_2^{(2)}K_2),
\label{fivetwo}
\end{equation}
with
\begin{equation}
J_1\equiv (\alpha b_3/b_2)\int_P d\zeta\biggl({\zeta^{n-1}-1\over 1-\zeta}
\biggr)
\int_0^\zeta {dy\over 1-y}{\ln(1+(1/t)\ln[(1-\zeta)(1-y)])\over (1+(1/t)
\ln[(1-\zeta
)(1-y)])^2},
\label{fivethreea}
\end{equation}
\begin{equation}
J_2\equiv -(\alpha b_3/b_2)\int_P d\zeta\biggl({\zeta^{n-1}-1\over 1-\zeta}
\biggr)
{\ln(1+(1/t)\ln(1-\zeta))\over (1+(1/t)\ln(1-\zeta))^2},
\label{fivethreeb}
\end{equation}
\begin{equation}
K_1\equiv -\int_Pd\zeta\biggl({\zeta^{n-1}-1\over 1-\zeta}\biggr)
\int_0^\zeta{dy
\over 1-y}{1\over (1+(1/t)\ln[(1-\zeta)(1-y)])^2},
\label{fivethreec}
\end{equation}
\begin{equation}
K_2\equiv \int_Pd\zeta\biggl({\zeta^{n-1}-1\over 1-\zeta}\biggr)
{1\over (1+(1/t)
\ln(1-\zeta))^2}\ .
\label{fivethreed}
\end{equation}
These integrals may be readily computed by separating the moment variable
dependence
 from the coupling constant dependence,
as in \cite{ref:two}, and using the contour ${\cal P}$ of fig. 1.
Care must be taken, however, regarding the singularity
structure of the integrals over the various pieces of the contour.
The semicircle contributions will now contain singular pieces of the
form $\ln^i\delta/\delta^j,\ i,j=0,1,..$
which will cancel with similar singularities coming from the straight-line
contributions. Let us consider, for example,
the integral $J_2$. Using the binomial expansion, we write
\begin{equation}
J_2=\sum_{m=1}^\infty {(1-n)_m\over m!}J_2^m\label{fivefour},
\end{equation}
with
\begin{equation}
J_2^m=-(\alpha b_3/b_2)\int_P d\zeta\zeta^{m-1}{\ln(1+(1/t)\ln\zeta)
\over(1+(1/t)\ln\zeta)^2}.
\label{fivefive}
\end{equation}
The semicircle contributions each contain an imaginary part, which cancels
upon addition, and a singular real part:
\begin{equation}
{1\over 2}J_2^m(\cap)+{1\over 2}J_2^m(\cup)
=-(b_3/b_2^2)t\zeta_t^m\biggl(-{2\ln(\delta/(t\zeta_t))\over \delta/\zeta_t}
-{2\over \delta/\zeta_t}+{m\pi^2\over 2}\biggr),
\label{fivesix}
\end{equation}
with $\zeta_t\equiv {\rm e}^{-t}$.
The straight-line integrations, on the other hand, give a combined
contribution:
\begin{eqnarray}
{1\over 2}J_2^m(+i\epsilon)+{1\over 2}J_2^m(-i\epsilon)&
=&-(b_3/b_2^2)t\zeta_t^m\biggl({2\ln(\delta/(t\zeta_t))\over \delta/\zeta_t}
+{2\over
 \delta/\zeta_t}\nonumber \\
& &-{1\over t\zeta_t^m}-{m\over 2}\int_{-\infty}^{mt}dx {\rm e}^x
\ln^2(|x|/(mt))
+m{\cal P}\int_{-\infty}^{mt}dx{{\rm e}^x\over x}\biggr)\ .
\label{fiveseven}
\end{eqnarray}
Upon addition we obtain the finite result
\begin{eqnarray}
J_2^m&=&{1\over 2}\biggl(J_2^m(+i\epsilon)+J_2^m(\cap)\biggr)+{1\over 2}
\biggl(J_2^m(-i\epsilon)+J_2^m(\cup)\biggr)\nonumber \\
&=&(b_3/b_2^2)(1-{\cal E}(mt))-(b_3/(2b_2^2))\Lambda(mt),
\label{fiveeight}
\end{eqnarray}
with
\begin{equation}
\Lambda(x)\equiv x{\rm e}^{-x}(\pi^2-\int_{-\infty}^x dy{\rm e}^y\ln^2(|y|/x))
\ .\label{fivenine}
\end{equation}
The function $\Lambda(x)$ is expressible in terms of
Incomplete Gamma functions, as
we will show later in this section.

Let us first give the results for the integrals $J_1,\ K_1,\ K_2$ appearing in
eqs.~(\ref{fivethreea})-(\ref{fivethreed}).
Performing the $y$-integration for the integral $J_1$ and expanding
in terms of the ``Pochhammer symbol'' $(1-n)_m$ as before, we have
\begin{equation}
J_1={b_3\over b_2^2}\sum_{m=1}^\infty {(1-n)_m\over m!}\biggl
\{J_{1;1}^m(t/2)-J_{1;1}
^m(t)
+J_{1;2}^m(t/2)-J_{1;2}^m(t)\biggr\},
\label{one}
\end{equation}
where we define
\begin{equation}
J_{1;1}^m(t)\equiv\int_P d\zeta \zeta^{m-1}{\ln(1+(1/t)\ln\zeta)
\over 1+(1/t)\ln
\zeta}
={1\over 2m}\Lambda(mt),
\label{two}
\end{equation}
and
\begin{equation}
J_{1;2}^m(t)\equiv\int_P d\zeta \zeta^{m-1}{1\over 1+(1/t)\ln\zeta}={1\over m}
{\cal E}(mt)\ .\label{three}
\end{equation}
Hence we obtain
\begin{equation}
J_1={b_3\over b_2^2}\sum_{m=1}^\infty {(1-n)_m\over m!m}\Biggl\{{1\over 2}
\biggl(
\Lambda
(mt/ 2)-\Lambda(mt)\biggr)+{\cal E}(mt/2)
-{\cal E}(mt)
\Biggr\}\ .\label{four}
\end{equation}
Working similarly for the integrals $K_1,\ K_2$, eqs. (5.4), (5.5), we find
\begin{equation}
K_1=-{1\over \alpha b_2}\sum_{m=1}^\infty {(1-n)_m\over m!m}
\{{\cal E}(
mt/2
)-{\cal E}(mt)\},
\label{five}
\end{equation}
and
\begin{equation}
K_2={1\over \alpha b_2}\sum_{m=1}^\infty {(1-n)_m\over m!}
\{{\cal E}(mt)-1\}\ .
\label{six}
\end{equation}
Putting everything together we find for the next-to-leading
exponent the expression
\begin{eqnarray}
E(n,\alpha)_{NL}&=&(\alpha b_3/b_2^2)\Biggl\{g_1^{(1)}\sum_{m=1}
^\infty
{(1-n)_m\over m!m}\biggl[{\cal E}(mt/2)-{\cal E}(mt)
+{1\over 2}(\Lambda(mt/2)-\Lambda(mt))\biggr] \nonumber \\
& &+g_2^{(1)}\sum_{m=1}^\infty{(1-n)_m\over m!}\biggl[{\cal E}(mt)-1+{1\over 2}
\Lambda(mt)\biggr]\Biggr\} \nonumber \\
& &+(\alpha/b_2)\Biggl\{-g_1^{(2)}\sum_{m=1}^\infty{(1-n)_m\over m!m}
\biggl[{\cal
 E}(mt/2)-{\cal E}(mt)\biggr] \nonumber \\
& &-g_2^{(2)}\sum_{m=1}^\infty
{(1-n)_m\over m!}\biggl[{\cal E}(mt)-1\biggr]\Biggr\}\ .
\label{fivefifteen}
\end{eqnarray}
To recover the resummed perturbative series, we have to perform
asymptotic expansions
 on the functions ${\cal E}(x)$,
and $\Lambda(x)$, eq.~(\ref{fivenine}).

As with the function ${\cal E}(x)\equiv x{\rm e}^{-x}Ei(x)$
(see eq.~(\ref{ieiofx})), we will now derive an
asymptotic formula for the function $\Lambda(x)$, eq.~(\ref{fivenine}).
The definition of this function is
\begin{equation}
\Lambda(x)\equiv x{\rm e}^{-x}(\pi^2-\int_{-\infty}^x dy{\rm e}^y\ln^2(|y|/x))
\simeq
-x{\rm e}^{-x}\biggl\{\gamma_2(x)-2\ln x\gamma_1(x)+\ln^2 x\gamma_0(x)
\biggr\}\ ,
\label{seven}
\end{equation}
where
\begin{equation}
\gamma_n(x)\equiv\int_0^x dy{\rm e}^y\ln^ny\ ,\label{sevena}
\end{equation}
the difference between the two expressions in eq.~(\ref{seven})\
being of order $(\ln {Q \over \Lambda}) {\Lambda \over Q}$.
The functions $\gamma_n(x)$ may be expressed in terms of Incomplete Gamma
functions \cite{ref:five}
\begin{equation}
\gamma_n(x)=\lim_{\epsilon\to 0^+}\biggl({\partial\over \partial\epsilon}
\biggr)
^n\int_0^x
dy{\rm e}^y y^\epsilon=\lim_{\epsilon\to 0^+}\biggl({\partial\over
\partial\epsilon}
\biggr)^n
x^{\epsilon+1}\Gamma(\epsilon+1)\gamma^*(\epsilon+1,-x)\ .\label{eight}
\end{equation}
Using the asymptotic expansion for the Incomplete Gamma function, or simply
integrating
eq.~(\ref{sevena})\ by parts, we obtain
\begin{equation}
\gamma_n(x)\simeq\lim_{\epsilon\to 0^+}\biggl({\partial\over \partial\epsilon}
\biggr)^n
x^\epsilon{\rm e}^x\Biggl\{1+\sum_{\rho=1}^{N'}{(-\epsilon)_\rho\over x^\rho}
\Biggr\},
\label{eighta}
\end{equation}
where $N'$ will be defined shortly.
Realizing that $(-\epsilon)_\rho$ is a polynomial
of $\rho$ degree in $\epsilon$,
 with
roots $\epsilon_j=j,\ j=0,...\rho-1$, we finally obtain
\begin{eqnarray}
\gamma_0(x)& \simeq &{\rm e}^x,\nonumber  \\
\gamma_1(x)& \simeq &{\rm e}^x\biggl[\ln x-{1\over x}\sum_{\rho=0}
^{N'-1}{\rho!\over x^\rho}\biggr],\nonumber \\
\gamma_2(x)& \simeq &{\rm e}^x\biggl[\ln^2 x-{2\ln x\over x}
\sum_{\rho=0}^{N'-1}{\rho!\over
 x^\rho}
+{2\over x}\sum_{\rho=1}^{N'-1}{[\Psi(\rho+1)
+\gamma]\rho!\over x^\rho}\biggr]\ .
\end{eqnarray}

Putting everything together we obtain the asymptotic expression
\begin{equation}
\Lambda(x)\simeq -2\sum_{\rho=1}^{N'}{[\Psi(\rho+1)+\gamma]\rho !\over x^\rho},
\label{fivesixteen}
\end{equation}
where $\Psi$ is the logarithmic derivative of the $\Gamma$-function.
As with $N\equiv\{N_i^L\}$ in eq.~(\ref{elpert}),
$N'$ depends on $x$ and different $N'$'s will
be determined for the different sums in eq.~(\ref{fivefifteen}).
A similar method was introduced in \cite{ref:two} for $E_L$.
Let us separate, therefore, the sums in eq.~(\ref{fivefifteen})\ and
derive optimum numbers of perturbative terms by direct optimization of
each sum separately.
Using eq.~(\ref{fivesixteen})\ along with the corresponding asymptotic formula
for ${\cal E}(x)$, \cite{ref:two}, we obtain the perturbative asymptotic
approximation:
\begin{eqnarray}
E(n,\alpha)_{NL}\simeq E(n,\alpha,N')_{NL} &=& (\alpha b_3/b_2^2)g_1^{(1)}
\sum_{\rho=0}^{N_1^{NL}}{\rho!2^\rho\over t^\rho}\sum_{m=1}^\infty
{(1-n)_m\over m!m^{\rho+1}}\nonumber \\
& &-(\alpha b_3/b_2^2)g_1^{(1)}\sum_{\rho=0}^{N_2^{NL}}{\rho!\over t^\rho}
\sum_{m=1}^\infty{(1-n)_m\over m!m^{\rho+1}}\nonumber \\
& &-(\alpha b_3/b_2^2)g_1^{(1)}\sum_{\rho=1}^{N_3^{NL}}
{\rho!2^\rho\over t^\rho}
[\Psi(\rho+1)+\gamma]\sum_{m=1}^\infty{(1-n)_m\over m!m^{\rho+1}} \nonumber \\
& &+(\alpha b_3/b_2^2)g_1^{(1)}\sum_{\rho=1}^{N_4^{NL}}{\rho!\over t^\rho}
[\Psi(\rho+1)+\gamma]\sum_{m=1}^\infty{(1-n)_m\over m!m^{\rho+1}} \nonumber \\
& &+(\alpha b_3/b_2^2)g_2^{(1)}\sum_{\rho=0}^{N_5^{NL}}{\rho!\over t^\rho}
\sum_{m=1}^\infty{(1-n)_m\over m!m^\rho} \nonumber \\
& &-(\alpha b_3/b_2^2)g_2^{(1)}\sum_{m=1}^\infty{(1-n)_m\over m!} \nonumber \\
& &-(\alpha b_3/b_2^2)g_2^{(1)}\sum_{\rho=1}^{N_6^{NL}}{\rho!\over t^\rho}[
\Psi(\rho+1)+\gamma]\sum_{m=1}^\infty{(1-n)_m\over m!m^\rho} \nonumber \\
& &-(\alpha/ b_2)g_1^{(2)}\sum_{\rho=0}^{N_1^{NL}}{\rho!2^\rho\over t^\rho}
\sum_{m=1}^\infty{(1-n)_m\over m!m^{\rho+1}} \nonumber \\
& &+(\alpha/b_2)g_1^{(2)}\sum_{\rho=0}^{N_2^{NL}}{\rho!\over t^\rho}
\sum_{m=1}^\infty{(1-n)_m\over m!m^{\rho+1}} \nonumber \\
& &-(\alpha/b_2)g_2^{(2)}\sum_{\rho=0}^{N_5^{NL}}{\rho!\over t^\rho}
\sum_{m=1}^\infty{(1-n)_m\over m!m^\rho} \nonumber \\
& &+(\alpha/b_2)g_2^{(2)}\sum_{m=1}^\infty{(1-n)_m\over m!}
\label{bigseries}
\end{eqnarray}

The set of numbers
$N'\equiv \{N_i^{NL}\},\ i=1,...6,$ is determined by direct optimization
of eqs.~(\ref{fivefifteen}), (\ref{bigseries}), as in the leading case.
Then, performing a Stirling approximation on the $n$-dependent
summations we observe that the fifth and sixth sums
contain at most one less power of $\ln n$ than
of coupling constant and hence we drop these for consistency.
Therefore we end up with the perturbative formula
\begin{equation}
E(n,\alpha,N')_{NL}=\sum_{i=1}^4\sum_{\rho=2}^{N_i^{NL}+1}\alpha^\rho
\sum_{j=0}^\rho s_{j,\rho}^{NL}(i)\ln^j n,
\label{pertcompact}
\end{equation}
where the perturbative coefficients are
\begin{eqnarray}
s_{j,\rho}^{NL}(1)&=&-[g_1^{(1)}(b_3/b_2^2)-g_1^{(2)}/b_2]b_2^{\rho-1}
(-1)^{\rho+j}{(\rho-1)!\over j!}2^{\rho-1}c_{\rho-j},\nonumber \\
s_{j,\rho}^{NL}(2)&=&[g_1^{(1)}(b_3/b_2^2)-g_1^{(2)}/b_2]b_2^{\rho-1}(-1)
^{\rho+j}
{(\rho-1)!\over j!}c_{\rho-j},\nonumber \\
s_{j,\rho}^{NL}(3)&=&g_1^{(1)}(b_3/b_2^2)b_2^{\rho-1}(-1)^{\rho+j}{(\rho-1)!
\over j!}2^{\rho-1}[\Psi(\rho)+\gamma]c_{\rho-j},\nonumber \\
s_{j,\rho}^{NL}(4)&=&-g_1^{(1)}(b_3/b_2^2)b_2^{\rho-1}(-1)^{\rho+j}{(\rho-1)!
\over j!}[\Psi(\rho)+\gamma]c_{\rho-j}.
\label{nlpertcoeff}
\end{eqnarray}
The perturbative exponent in moment space, resumming all large perturbative
corrections, becomes, after eqs.~(\ref{elpert}), (\ref{elcoeff}),
(\ref{pertcompact}), (\ref{nlpertcoeff}) :
\begin{equation}
E(n,\alpha, {\cal N})=E(n,\alpha,N)_L+E(n,\alpha,N')_{NL},
\label{exptot}
\end{equation}
with ${\cal N}={\cal N}(t)\equiv \{N;N'\}=\{N_1^L,N_2^L,N_3^L;N_1^{NL},
N_2^{NL},N_3^{NL},N_4^{NL}\}$, all numbers being functions of $t$ obeying
the relations
\begin{equation}
N_1^L(t)=N_2^L(t/2),\ N_1^{NL}(t)=N_2^{NL}(t/2)=N_3^L(t),\ N_3^{NL}(t)
=N_4^{NL}(t/2)\ ,
\label{nrel}
\end{equation}
which give a total of three independent functions. The resummed hard parts
in momentum space can now be obtained, given the exponent
eq.~(\ref{exptot}). In the perturbative regime, a closed formula for the
hard parts  giving all large perturbative corrections is \cite{ref:four}:
\begin{equation}
I(z,\alpha)=\delta(1-z)-\Biggl[{{\rm e}^{E\left({1\over 1-z},\alpha, {\cal N}
\right)}\over \pi(1-z)}\Gamma\left(1+P_1\left({1\over 1-z},\alpha,{\cal N}
\right)\right)\sin\left(\pi P_1\left({1\over 1-z},\alpha,{\cal N}\right)\right)
\Biggr]_+\ ,
\label{hardpart}
\end{equation}
with
\begin{equation}
P_1(n,\alpha,{\cal N})\equiv {\partial\over \partial\ln n}E(n,\alpha,{\cal N})
\ .
\label{pone}
\end{equation}
The range of validity of eq.~(\ref{hardpart})\ is $2\alpha b_2\ln(1/(1-z))<1$
and in fact, numerical calculations in \cite{ref:two} suggest that this
upper limit is in good agreement with
the peak position of the exact leading exponent,
$\alpha b_2\ln(1/(1-z_p^L))\simeq 0.6$.
For larger $1/(1-z)$, the higher-twist inherent in the resummation formula,
takes over, causing the perturbative approximation to break down.
In the next section we will determine the behavior of the higher twist
analytically, as a large-$n$ asymptotic approximation in a regime
complementary to the perturbative one, namely for $\alpha b_2\ln(1/(1-z))>1$.
In section 5 we will present numerical results for $E(n,\alpha)_{L/NL}$
and $E(n,\alpha,N/N')_{L/NL}$ along with the sets of numbers $N$ and $N'$,
for various values of $\alpha$.

\section{Behavior of the exponent in the non-perturbative regime}

Before proceeding to numerical results, it is important to understand
analytically
the behavior of the exponent for asymptotically large values of $n$
-beyond the perturbative regime. This will reveal both the behavior of
the higher twist inherent in our resummation formalism, as well as
that of the cross section itself at the very edge of phase space
$z\to 1$, or equivalently, $n\to \infty$. It will also confirm and
extend the corresponding numerical calculations, which are quite
sensitive at this range. In \cite{ref:two} a heuristic argument
for the behavior of $E(n,\alpha)_L$ was given, which was verified
by numerical results showing that
$$\lim_{n\to\infty}E(n,\alpha)_L=-\infty\ .$$
Here we will elaborate on this argument, developing asymptotic formulas
for both $E(n,\alpha)_L$ and $E(n,\alpha)_{NL}$ in the non-perturbative
range  $1<\alpha b_2\ln n<\infty$.
These formulas will be checked against numerical results in section 5.

Let us begin with the leading exponent, eq.~(\ref{elintegrals}). We
will concentrate on the integral $I(n,t)$, eq.~(\ref{ienti}). After a
change of variables $\zeta\to 1-\zeta$, it reads
\begin{equation}
I(n,t)=t\int_P{d\zeta\over \zeta}W(n,\zeta)\ln(1+(1/t)\ln\zeta),
\label{ientiw}
\end{equation}
with the $n$-dependent weight given by
\begin{equation}
W(n,\zeta)\equiv[(1-\zeta)^{n-1}-1]={\rm e}^{-(n-1)\ln\left({1\over 1-\zeta}
\right)}-1\ .
\label{weight}
\end{equation}
The branch-point singularity of the above integrand is at
$\zeta_t={\rm e}^{-t}$. On the other hand, the weight approaches the
value $-1$ when\footnote{This value is the source of divergent behavior
as $n\to\infty$.}
\begin{equation}
\zeta>1-{\rm e}^{-{1\over n-1}}\equiv n_0(n)\ .
\label{enzerro}
\end{equation}
As $n$ varies, we may have either $\zeta_t<n_0(n)$ or $\zeta_t>n_0(n)$.
The former inequality is satisfied whenever
\begin{equation}
n<1+{1\over \ln\left({1\over 1-{\rm e}^{-t}}\right)}\equiv \zeta_1(t).
\label{former}
\end{equation}
In fact, for reasonably small values of $\alpha$,
$\zeta_1(t)\simeq \zeta_t^{-1}={\rm e}^t=Q^2/\Lambda^2$
and hence eq.~(\ref{former})\ is equivalent to
$\alpha b_2\ln n<1$, which we recognize as our perturbative regime for
this particular integral. As before, some of the integrals in $E_L$
and $E_{NL}$ contain the scale $t/2$, but again we take the intersection
of the regions defined by $\zeta_1(t),\ \zeta_1(t/2)$ to define the
narrower non-perturbative regime $n>\zeta_1(t)\simeq {\rm e}^t$.
This leaves a rather uninteresting
intermediate region $Q/\Lambda<n<Q^2/\Lambda^2$, where
various scales impose different asymptotic behaviors to different
pieces of the exponent.

Therefore the non-perturbative asymptotics are described by the conditions
\begin{equation}
n>\zeta_1(t)\simeq {\rm e}^t={Q^2\over \Lambda^2},
\ W(n,\zeta)\simeq -1,\ \zeta>n_0(n)\simeq {1\over n},
\label{conditions}
\end{equation}
where the asymptotic behavior of $I(n,t)$ is given by
\begin{equation}
I(n,t)_>\simeq-t\int_{{\cal P}_0}{d\zeta\over \zeta}\ln(1+(1/t)\ln\zeta),
\label{ientiapp}
\end{equation}
with ${\cal P}_0$ the principal-value contour proper, fig. 1b, but starting
fron $n_0(n)$, rather than 0.
We have computed the integral on this contour in the appendix. The result is
\begin{equation}
I(n,t)_>
\simeq t\biggl\{-tL(n,t)\ln L(n,t)+\ln(1/n_0(n))\biggr\}
\equiv tG(n,t),
\label{ientiasym}
\end{equation}
with
\vspace{-7mm}
\begin{eqnarray}
&n_0(n)=1-{\rm e}^{-{1\over n-1}}\simeq {1\over n},\nonumber \\
&L(n,t)\equiv {1\over t}\ln\left({1\over n_0(n)}\right)-1\simeq{\ln n\over t}
-1,\nonumber \\
&n>\zeta_1(t)=1+{1\over \ln\left({1\over 1-{\rm e}^{-t}}\right)}\simeq
{\rm e}^t
\ .
\label{condi}
\end{eqnarray}

Let us now digress for a while, to complete the discussion in section 2
regarding the definition of the perturbative regime. We will confine
our discussion to the integral $I(n,t)$ as given by eq.~(\ref{ienti}).
In section 2, we showed that a general expression for this integral is

\begin{equation}
I(n,t)=-\sum_{m=1}^\infty{(1-n)_m\over m!m^2}{\cal E}(mt)\ .
\label{gexpression}
\end{equation}
We also showed that for
\begin{equation}
t>1,\ \ 1\ll n< {\rm e}^t,
\label{condition}
\end{equation}
we obtain the perturbative approximation
\begin{equation}
I(n,t)\simeq -\sum_{\rho=0}^N{\rho!\over t^\rho}(-1)^{\rho+1}\sum_{
j=0}^{\rho+2}
c_{\rho+2-j}{(-1)^j\over j!}\ln^j n\equiv I(n,t,N)\ .
\label{ienten}
\end{equation}
Even though the second condition in eq.~(\ref{condition}) is the one
which suppresses the logarithms and hence makes an asymptotic expansion
possible in $I(n,t,N)$, only the
first condition in eq.~(\ref{condition}) was used to derive this approximation.
Hence, the question might arise whether the perturbative approximation
(\ref{ienten}) is valid independently of the second condition in
(\ref{condition}),
which defines the perturbative regime in $n$.
We now see from eqs.~(\ref{conditions}), (\ref{ientiasym}),
(\ref{condi}), however, that for $n>\zeta_1(t)\simeq {\rm e}^t$, the
integral $I(n,t)$ is approximated by $I(n,t)_>$ which is clearly
a non-perturbative expression. Furthermore, as we will see in the rest of
this section, expressions like $I(n,t,N)\ {\rm and} \ I(n,t)_>$
are numerically very different from one another, for $n>{\rm e}^t$.
\footnote{$I(n,t)_>$, on the other hand, is not even defined for
$n<{\rm e}^t$.}  Thus, when $t>1$,
the perturbative expression $I(n,t,N)$ approximates $I(n,t)$ for $n<{\rm e}^t$
while the non-perturbative expression $I(n,t)_>$ approximates
$I(n,t)$ for $n>{\rm e}^t$.   This justifies the
characterization of the  corresponding  $n$-ranges
as perturbative and non-perturbative, respectively.

Returning  now to our calculation, we find for $I_2(n,t)$, eq.~(\ref{itwo}),
\begin{equation}
I_2(n,t)_>\simeq t\ln L(n,t)\equiv H(n,t),
\label{itwoasym}
\end{equation}
with $L(n,t)$ given by eq.~(\ref{condi}).
Our leading exponent is then given, in this region, by
\begin{equation}
E(n,\alpha)_{L>}=\alpha g_1^{(1)}[2I(n,t/2)_>-I(n,t)_>]-
\alpha g_2^{(1)}I_2(n,t)_>\ .
\label{elinsert}
\end{equation}
 From eqs.~(\ref{ientiasym}-\ref{elinsert})\ we find that, when
$n>{\rm e}^{{\rm e}t}$,
$$G(n,t)\simeq-\ln n\ln\left({\ln n\over t}\right),$$
and hence
\begin{equation}
E(n,\alpha)_{L>}\simeq -{g_1^{(1)}\over b_2}\ln 2\ln n-{g_2^{(1)}\over
b_2}\ln\left({\ln n\over t}\right)\ .
\label{elasym}
\end{equation}
The first term above dominates the behavior as a logarithmic divergence
in $n$, with a negative coefficient. This point was also made in
\cite{ref:two}, but the precise asymptotic behavior,
eqs.~(\ref{ientiasym})-(\ref{itwoasym}),
valid in almost all of the non-perturbative range $n>\zeta_1(t)$,
was not given.

We can similarly find the large-$n$ asymptotics of the
next-to-leading exponent,
\begin{equation}
E(n,\alpha)_{NL}=\alpha[g_1^{(1)}J_1(n,t)-g_2^{(1)}J_2(n,t)]
+\alpha^2[g_1^{(2)}K_1(n,t)-g_2^{(2)}K_2(n,t)]\ .
\label{insertp}
\end{equation}
Concentrating on the integrals $J_1$ and $K_1$, eq.~(\ref{fivetwo}), and
using the results of the appendix, we can write
\begin{equation}
J_1(n,t)_>\simeq {b_3\over b_2^2}\biggl\{F(n,t/2)-F(n,t)
+H(n,t/2)-H(n,t)\biggr\},
\label{joneasym}
\end{equation}
with $H(n,t)$ defined in eq.~(\ref{itwoasym}), where
\begin{equation}
F(n,t)={t\over 2}\biggl\{\ln^2(L(n,t))-\pi^2\biggr\},
\label{fnt}
\end{equation}
and
\begin{equation}
K_1(n,t)_>\simeq -t\biggl\{H(n,t/2)-H(n,t)\biggr\}\ .
\label{koneasym}
\end{equation}
Similarly, for $J_2$ and $K_2$ we find
\begin{equation}
J_2(n,t)_>\simeq -{b_3\over b_2^2}P(n,t),
\label{jtwoasym}
\end{equation}
where
\begin{equation}
P(n,t)={1\over L(n,t)}\ln L(n,t)+{1\over L(n,t)}+1,
\label{pnt}
\end{equation}
and
\begin{equation}
K_2(n,t)_>\simeq t\biggl\{{1\over L(n,t)}+1\biggr\}\equiv Q(n,t)\ .
\label{ktwoasym}
\end{equation}
Notice from the above expressions that the quantities $J_2$, $K_2$,
which were subleading in the perturbative regime, remain subleading
in the non-perturbative regime as well. In fact, the dominant large-$n$
behavior in $E(n,\alpha)_{NL}$ comes from the function
$F(n,t)$, eq.~(\ref{fnt}). For values
\begin{equation}
n>1+{1\over \ln\left({1\over 1-{\rm e}^{-t[1+{\rm e}^\pi]}}\right)}
\simeq{\rm e}^{t[1+{\rm e}^\pi]}\ ,
\label{extreme}
\end{equation}
the next-to-leading exponent is dominated by
\begin{equation}
E(n,\alpha)_{NL>}\simeq \alpha
g_1^{(1)}{b_3\over b_2^2}\biggl\{{t\over 4}\ln^2\left
({2\ln n\over t}\right)-{t\over 2}\ln^2\left({\ln n\over t}\right)\biggr\}
\simeq-g_1^{(1)}{b_3\over 4b_2^3}\ln^2\left({\ln n\over t}\right)\ .
\label{enlasym}
\end{equation}
Therefore, $E(n,\alpha)_{NL>}$ also tends to $-\infty$,
though more slowly than $E(n,\alpha)_{L>}$. This only happens, however,
for huge values of $n$,
eq.~(\ref{extreme}), quite inaccessible even
to numerical calculations. As we shall see
in section 5, within the numerically accessible range of $n$, the
singularity manifests itself through the presence of $\pi^2$, eq.~(\ref{fnt}),
which dominates, making $E(n,\alpha)_{NL}$ {\it look like} it is tending to
$+\infty$ within that range. This is a demonstration of the importance
of the above asymptotic formulas where numerical calculations break down.

\section{Numerical results for the Exponent and the Hard Part}

In this section we present numerical results for the exponent of the
resummation formula as well as the $q{\bar q}$ hard part, at various values
of the kinematic variables, and including the large perturbative
corrections. In our opinion, these results exhaust our knowledge
of the perturbative phase of the theory. The various constants appearing
throughout this work, are
\begin{eqnarray}
&g_1^{(1)}=2C_F,\ g_2^{(1)}=-{3\over 2}C_F,\ g_1^{(2)}=C_F\left[C_A
\left({67\over 18}-{\pi^2\over 6}\right)-{5n_f\over 9}\right],\nonumber \\
&b_2=(11C_A-2n_f)/12,\ b_3=(34C_A^2-(10C_A+6C_F)n_f)/48.
\label{constants}
\end{eqnarray}
For the rest of this paper, we will use the values \cite{ref:two}
$\Lambda=0.2{\rm GeV}$, $n_f=4$ and $Q=5,\ 10,\ 90{\rm GeV}$, which
correspond to $\alpha=0.075,\ 0.061,\ 0.039$ respectively.

\subsection{The resummed exponent}

Let us first present numerical results for the exponent in moment space.
Approximate conclusions in momentum space may be reached through the
correspondence $n\leftrightarrow 1/(1-z)$.
As in \cite{ref:two}, let us first give the optimum numbers of asymptotic
terms for eqs.~(\ref{bigseries}), (\ref{pertcompact}), for the above values
of $Q$, computed at the intermediate value $n=30$.\footnote{
As with the leading exponent, there is very
little dependence of these numbers on $n$, when the latter is well
within the perturbative regime. Strictly speaking,
for $Q= 5{\rm GeV}$,  $n=30$ reaches the limit $Q/\Lambda$
for $N_1^L,\ N_1^{NL},\ N_3^{NL}$.
However, these numbers are unchanged for lower
values of $n$, as well.}
As we have noted before, the set ${\cal N}$ depends on
$t=1/ (\alpha b_2)=\ln(Q^2/\Lambda^2)$.
\begin{center}
\begin{tabular}{||cccccccc||}
\multicolumn{8}{c} {TABLE 1} \\ \hline
\multicolumn{8}{|c|}{Optimum numbers of asymptotic terms for $E(
n,\alpha)_L, E(n,\alpha)_{NL}$} \\ \hline\hline
$\alpha(t)$ & $N_1^L$ & $N_2^L$ & $N_3^L$ & $N_1^{NL}$ & $N_2^{NL}$ & $N_3^{N
L}$ & $N_4^{NL}$ \\  \cline{1-8}
0.075(6.4)  & 1 &  5 &  5 & 1 &  5 & 1 &  5 \\ \hline
0.061(7.8)  & 2 &  6 &  6 & 2 &  6 & 1 &  6 \\ \hline
0.039(12.2) & 4 & 12 & 11 & 4 & 11 & 4 & 11 \\ \hline
\end{tabular}
\end{center}

Given these values we may compute the perturbative exponent
as given in eqs.~(\ref{exptot}), (\ref{elpert}), (\ref{elcoeff}),
(\ref{pertcompact}), (\ref{nlpertcoeff})\  and
compare it with the exact exponent $E(n,\alpha)=E(n,\alpha)_L+
E(n,\alpha)_{NL}$. The latter is computed numerically on the
box contour \cite{ref:two} ${\bar P}$, fig. 1c.
In fig. 2 we have plotted the result
of these comparisons for the above values of $\alpha$.
For completeness we have also included
comparisons of the leading exponents alone.
Notice that we expect the peak of the exact exponents to grow
as $\alpha$ gets smaller. This is due to the fact that the
higher-twist tends to suppress the exponent, hence, the smaller the
$\alpha$, the less the suppression.
Notice also that the exact and perturbative curves are in excellent
agreement near the peak of the exact curves. We will comment on this
in some detail later.
Once past the peak of $E(n,\alpha)$, the higher-twist completely
dominates the exponent.

Using the results of sec. 4 we may
compare the exact exponents with their large-$n$ asymptotic
expressions, eqs.~(\ref{elasym}), (\ref{enlasym}) in the non-perturbative
regime. This is done
in fig. 3 for various values of $\alpha$, within the maximum
range of $n$ attainable before we run into round-off errors
for the exact exponents. Note that the agreement between the exact curves
and the analytical approximations improves as $n$ increases.

In fig. 4a we have plotted separately
$E(n,\alpha)_{NL}$ and the corresponding perturbative approximation
for various values of $Q$. We see that both curves have the
same shape
for all values of $Q$ used, but that at low $Q$-values
numerical agreement is only fair.
We can trace this to the determination of $N'$ and
$N_3^{NL}$ in particular, see eqs.~(\ref{fivefifteen}), (\ref{bigseries}),
and table 1. Using the exact expression for $\Lambda(x)$
as given by the first equality of eq.~(\ref{seven}), we see
that the $\pi^2$-factor is obviously of higher-twist origin and
can not be expressed as a polynomial in $\alpha$. However,
its effect on the third sum of eq.~(\ref{fivefifteen})\ may be isolated.
The equation determining $N_3^{NL}$ may be written
\begin{equation}
{1\over 2}\sum_{m=1}^{n-1}{(1-n)_m\over m!m}\Lambda(mt/2)
\simeq -\sum_{\rho=1}^{N_3^{NL}}{\rho!2^\rho\over t^\rho}
[\Psi(\rho+1)+\gamma]\sum_{m=1}^{n-1}{(1-n)_m\over m!m^{1+\rho}}\ .
\label{comparenl}
\end{equation}
The LHS gives, for $Q=10{\rm GeV}$, the values $\{1.798,0.363\}$
for $n=\{10,30\}$ respectively. At the same time the {\it best}
approximation on the RHS is given for $N_3^{NL}=1$ and equals the
minimum possible values
$\{1.219,2.212\}$ respectively, which is admittedly not a  very
good perturbative approximation, with an error relative to the
perturbative value of $\{47\%,83\%\}$
respectively.
 If, on the other hand,
we subtract the effect of the $\pi^2$-term in the LHS of eq.~(\ref{comparenl})
that enters the definition of $\Lambda(x)$ in eq.~(\ref{fivenine}),
which may be computed analytically to be
$${\pi^2t\over 4}[(1-{\rm e}^{-t/2})^{n-1}-1]\ ,$$
we obtain the values $\{5.008,8.923\}$. These can be approximated
by the perturbative expression for $N_3^{NL}=3-4$ yielding
the values $\{5.568,8.395\}$, in very good agreement with
the previous ones, with
an error of only $\{10\%,6\%\}$. For values of $Q$ near the Z-mass,
on the other hand, numerical agreement between the two curves
is excellent in the perturbative regime.
To summarize, the accuracy of the perturbative
approximation for $E(n,\alpha)_{NL}$ is fair for low $Q$ because
the higher-twist is much more important, and is suppressed more
slowly with increasing $Q$, than in the integrals giving the leading
exponent, $E(n,\alpha)_L$.

On the other hand, in fig. 4b we have plotted
separately $E(n,\alpha)_{NL}$ in the nonperturbative regime
and have compared it with the corresponding asymptotic approximation
$E(n,\alpha)_{NL>}$, up to the maximum values of $n$ where
a reliable numerical calculation of the former is possible.
Notice the excellent agreement as well as the fact that these
quantities have a positive increasing behavior in this range,
something exhibited by the dominance of the $\pi^2$-term in
eq.~(\ref{fnt}), as remarked in sec. 4. Of course, for even larger
values of $n$, the next-to-leading exponent tends to
$-\infty$ as suggested by eqs.~(\ref{extreme}), (\ref{enlasym}).
\footnote{These values are inaccesible to an exact numerical calculation but
are very easily handled numerically through the
approximate expressions, eqs.~(\ref{insertp}-\ref{ktwoasym}).}

Let us now return to fig. 2 and discuss in more detail the
peak positions of the exact exponents. In \cite{ref:two} it was found that
the peak of  $E(n,\alpha)_L$ is pretty close to the perturbative
boundary $n\simeq Q/\Lambda$. Adding $E(n,\alpha)_{NL}$ decreases
both the magnitude and the position of the peak, as seen in the above
figures. Let us therefore
define some characteristic points in these  curves which will describe,
within a controlled arbitrariness, the interface
between the perturbative and non-perturbative regime and a corresponding
theoretical uncertainty in the resummed perturbative cross section.
One such point that we have already mentioned several times,
${\rm e}^{t/2}=Q/\Lambda$,
separates the perturbative from the exact analytical expression
for the exponent. This, from an analytical point of view, is clearly
the boundary between the perturbative and non-perturbative regimes.
A second point, already mentioned in \cite{ref:two},
is the peak position of the exponent, $\zeta_p:\ E(\zeta_p,\alpha)={\rm max}$.
In fact we may define separately the peak positions for the leading
exponent\footnote{This point was denoted by $n_1$ in \cite{ref:two}.},
$\zeta_p^L$ and the full one, $\zeta_p$.
The table that follows gives these numbers as a function of $Q$.
\begin{center}
\begin{tabular}{||c@{\hspace {1in}}c@{\hspace {1.1in}}c@{\hspace {1.1in}}l||}
\multicolumn{4}{c} {TABLE 2} \\ \hline
\multicolumn{4} {|c|} {Boundaries in $n$ - space parametrizing the
perturbative/nonperturbative interface} \\ \hline\hline
$Q(GeV)$ & $Q/\Lambda$ & $\zeta_p^L$ & $\zeta_p$ \\ \cline{1-4}
 5       &    25       &     37      &    14 \\ \hline
10       &    50       &    100      &    37 \\ \hline
90       &   450       &   1875      &   570 \\ \hline
\end{tabular}
\end{center}

Notice in the above table that inclusion of $E_{NL}$ creates good
agreement between the order of $Q/\Lambda$ and $\zeta_p(Q)$. That is, while
$2\alpha b_2\ln \zeta_p^L$ is close to unity with good accuracy,
 ranging from 1.12 ($Q=5{\rm GeV}$) to 1.24 ($Q=90{\rm GeV}$), inclusion of
the next-to-leading exponent moves the peak so that
$2\alpha b_2\ln\zeta_p(5;10;90{\rm GeV})=0.82;\ 0.92;\ 1.03$, is close
to 1. In fact these results strongly support
the view that the peak of the exponent occurs at the limit of
validity of perturbation theory and {\it approximately coincides}
with the analytical result, namely $Q/\Lambda$.
The fact that we have two different scales in the problem, $t$
and $t/2$, is reflected in the minor differences between $Q/\Lambda$
and $\zeta_p$ in table 2. This difference may be used to provide
a small theoretical uncertainty in the resummed perturbative
hard part but, for the purpose of this work we will consider
${\rm e}^{t/2}=Q/\Lambda$ to be the boundary of the perturbative
regime in moment space. In momentum space, this converts  into a perturbative
regime given by
\begin{equation}
0<z<1-{\Lambda\over Q}\ ,
\label{pertz}
\end{equation}
and this is what we will use for defining all resummed perturbative
quantities below.

\subsection{The resummed hard part}

A formula for the hard part of the $q{\bar q}$ cross section,
 which certainly contains all the
perturbative corrections that eq.~(\ref{hardpart})\ does, but where the
higher-twist is not separated, may be obtained by ``extending"\footnote{
Strictly speaking, the inversion of the Mellin transform \cite{ref:four}, was
performed only for the perturbative exponent. The corresponding
formula for the hard parts, involving power-counting arguments
for all large perturbative corrections, simplifies into eq.~(\ref{hardpart})\
only for the polynomial approximation of the exponent as powers
of $\alpha$ and $\ln n$. }
eq.~(\ref{hardpart})\  throughout the whole range of $z$, by numerically
calculating both the full exponent $E(1/(1-z),\alpha)$ and
its derivative
$$P_1\left({1\over 1-z},\alpha\right)={\partial\over \partial \ln(1/(1-z))}
E\left({1\over 1-z},\alpha\right),$$
on the ``box" contour ${\bar P}$, fig. 1:
\begin{equation}
I(z)=\delta(1-z)-\Biggl[{{\rm e}^{E\left({1\over 1-z},\alpha\right)}\over
\pi(1-z)}\Gamma\left(1+P_1\left({1\over 1-z},\alpha\right)\right){\rm sin}
\left(\pi P_1\left({1\over 1-z},\alpha\right)\right)\Biggr]_+\ .
\label{hardpartf}
\end{equation}
As is discussed in \cite{ref:four}, in the physical cross section
we may perform an integration by parts
 and explicitly
get rid of the plus-distributions as well as the $\delta$-function
in the above formula.
Then, the relevant momentum-dependent quantity contributing
directly to the cross section as the perturbative hard part is
\begin{equation}
{\cal H}(z,\alpha)=\int_0^z dz'{{\rm e}^{E\left({1\over 1-z'},
\alpha\right)}
\over \pi (1-z')}\Gamma\left(1+P_1\left({1\over 1-z'},
\alpha\right)\right)
{\rm sin}\left(\pi P_1\left({1\over 1-z'},\alpha
\right)\right).
\label{chardpart}
\end{equation}
On the other hand, from the previous section,
and using the results in  tables 1 and 2,
we can now
find the resummed perturbative
hard part by separating the higher twist.
Since the latter has the effect of reducing the exponent till,
as $n\to\infty$, it goes to $-\infty$, we may remove its effect by
setting the {\it perturbative}
hard part equal to zero in the
non-perturbative regime $1-\Lambda/Q\le z\le 1$. We thus arrive at
the following perturbative formula:
\begin{eqnarray}
{\cal H}(z,\alpha, {\cal N})=&\Theta\left(1-{\Lambda\over Q}-z\right)
{\displaystyle \int_0^z}dz'{\displaystyle {{\rm e}^{E\left({1\over 1-z'},
\alpha,{\cal N}
\right)} \over \pi(1-z')}}\Gamma\left(1+P_1\left({1\over 1-z'},\alpha,
{\cal N}\right)\right)\nonumber\\
&\times{\rm sin}\left(\pi P_1\left({1\over 1-z'},\alpha,{\cal N}\right)\right).
\label{hardpartpert}
\end{eqnarray}

To avoid repetition, let us define the
notation that, for any resummed quantity of interest $R$,
$R({\cal N})_L$ stands for this quantity
computed using eq.~(\ref{hardpartpert})\ with $E({\cal N})_L$
only, $R({\cal N})$ using the same equation but with the full
$E({\cal N)})$ and similarly for $R_L$ and $R$, but
using eq.~(\ref{chardpart})\
instead.
We will now present numerically the hard part computed according to these
various definitions.

In fig. 5 we have plotted the functions ${\cal H}(z,\alpha)_L$,
${\cal H}(z,\alpha)$ as well as their perturbative
counterparts.
Some comments on these curves are in order.
As a general observation, the hard part calculated with the
total exponent agrees very well with its perturbative approximation in
almost all of the $z$-range and the agreement improves with
increasing $Q$. In particular, near the $Z$-mass the
agreement is excellent.
In the neighbourhood of the non-perturbative regime, $z>1-\Lambda/Q$,
we observe some interesting behavior, naturally more pronounced at
low $Q$-values.
Let us first concentrate on ${\cal H}(z,\alpha,{\cal N})_L$, computed
from eq.~(\ref{hardpartpert}). Going for simplicity to $n$-space,
we observe that for any value of $Q$, the polynomial
$P_1^L(n,\alpha,{\cal N})$ is a positive increasing function of
$n$ in the perturbative regime
but may be of order unity for sufficiently large values of $Q$.
This was actually anticipated in \cite{ref:four}, and does not
contradict the power-counting of $\alpha$ and $\ln n$ developed
in that reference, which established that $P_1$ contains at
most equal powers of these quantities and hence is anticipated
to be $ \le 1$ in the range $2\alpha b_2\ln n<1$. In fact this expectation
is numerically accurate for the {\it normalized} series
$P_1/g_1^{(1)}$, for all values of $Q$. The somewhat large (and
arbitrary from a resummation point of view) value of the
color factor $g_1^{(1)}=2C_F=8/3$, which is automatically included in
the resummation as an overall multiplicative factor of
$E(n,\alpha,{\cal N})_L$, as it should, can make $P_1$ somewhat larger.
To be concrete, for $Q=5,\ 10,\ 90{\rm GeV}$,
$P_1(n=Q/\Lambda)_L= 0.79,\ 1.67,\ 2.74$
respectively while the corresponding normalized series
has the values 0.30,\ 0.62,\ 1.02,
in agreement with the power-counting estimate.
For the values $P_1(n,\alpha,{\cal N})_L>1$ at corresponding
values of $Q$,  the sinusoidal dependence
of the hard part will produce a decreasing behavior which may
even become oscillatory for sufficiently large $Q$. The decreasing
behavior is shown in fig. 5b, whereas the oscillatory behavior occurs
for $Q=90{\rm GeV}$ but is not shown. These effects are obviously concentrated
in the neighborhood of $z\simeq 1-\Lambda/Q$ in momentum space, but are
still part of the perturbative regime due primarily
to the overall color factor.

Similar remarks apply to  the ``exact" curves, produced through the
formula (\ref{chardpart}), where the sign of the integrand
will become negative whenever $P_1>1$ or $P_1<0$.
We note that, at
$Q=5{\rm GeV}$ both $P_1^L$ and $P_1$ are
less than 1 in the entire perturbative regime,
at $Q=10{\rm GeV}$ $P_1^L>1$ for $15<n<38$ while $P_1<1$ in the entire
perturbative regime and at $Q=90{\rm GeV}$ $P_1^L;\ P_1>1$
for  $39<n<935$ and $27<n<319$ respectively.
These ranges will create oscillations in the
corresponding curves for the hard part, as is shown in fig. 5b for
${\cal H}(z,\alpha)_L$. As we tend towards the peak
position of the
corresponding exponent, table 1, $P_1^L; P_1$ will decrease again, change
sign as they cross the peak position, and become  negative
in the non-perturbative regime. There is no
danger from the poles of the $\Gamma$-function when $P_1$ becomes a
negative integer, because ${\rm sin}(\pi P_1)$, that multiplies it,
cancels these poles.
The corresponding curves
will  decrease thereafter, reaching a finite value at $z=1$
since, as we have seen in sec. 4, the exponent of
the resummation formula (\ref{chardpart}) tends to
$-\infty$ as $z\to 1$. We indeed find
numerically that, for $Q=5,\ 10,\ 90 {\rm GeV}$,
${\cal H}(z\simeq 1,\alpha)_L\simeq
-2.17;-10.85;-644.0$
respectively, while
${\cal H}(z\simeq 1,\alpha)\simeq -0.72;-2.23;-151.4$.
\footnote{The values quoted for ${\cal H}$ are taken at the highest
value of $z$, equal to $0.999999$, before machine round-off errors are
encountered by the program.}
The corresponding
curves for the cross section will be bounded for values of $\tau$ near the
edge of phase space $\tau\simeq 1$.

\section{Conclusions}

In this paper we have completed the principal-value
resummation  of the Drell Yan cross section in the DIS
scheme, by including the next-to-leading exponent to previously
existing results \cite{ref:two}. The resulting exponent $E=E_L+E_{NL}$
includes the large perturbative corrections appearing as plus-distributions
in the perturbative calculation of the cross section and produces
a resummed hard part \cite{ref:four} that is finite throughout
the entire kinematic range and independent of IR cutoffs.
This principal-value exponent $E$ may be approximated by a perturbative
expression, $E({\cal N})$,  obtained in moment space
as an asymptotic series  in the perturbative
regime $ 1<<n<Q/\Lambda$, and by a non-perturbative expression, $E(n)_>$
when $n>Q^2/\Lambda^2$. The former contains a finite number ${\cal N}(t)$ of
perturbative terms,
which depends on three independent functions of
$t\equiv \ln(Q^2/\Lambda^2)$, while the latter tends to $-\infty$
as $n\to\infty$. This makes the hard part a finite function of $z$
and the resummed cross section a bounded function of $\tau$.

The accuracy of the perturbative approximation of the exponent
improves with $Q$, as expected, being good at fixed-target
values and excellent near the Z-mass.
$E_{NL}({\cal N})$ approximates fairly
its exact version at low $Q$-values,
due to large higher-twist contributions, and very well near the Z-mass.
In short, $E_{NL}$ is numerically significant,
as expected, in all its versions and all regions.

The  hard part for the Drell Yan cross section, ${\cal H}(z,\alpha)$,
whose general form was given in \cite{ref:four},
is reproduced accurately in the perturbative regime by its perturbative
approximation ${\cal H}(z,\alpha,{\cal N})$ except near the edge
$z=1-\Lambda/Q$, where both expressions peak and then oscillate
(reaching negative values) due to higher-twist  and/or
large color factors. As before, agreement improves
substantially with increasing $Q$, making the theoretical
uncertainty in the resummation, if defined roughly as the difference between
the above two quantities,
a decreasing function of Q.

We are now in a position to input this resummed hard part,
with the properties described above, directly to a convolution with
the non-perturbative parton flux for various experimental situations,
to obtain
the corresponding resummed cross section. Due to the multitude of
experimental cases to be covered, the associated numerical complication
of the corresponding calculations, the need for an accurate
definition of the uncertainty in the resummation
and the essentially phenomenological
character of   these matters, we reserve them for
a subsequent publication.
\vskip 1in
\centerline{ACKNOWLEDGMENTS}

We would like to thank Ed Berger for very informative discussions
and George Sterman for his generous and constant support, in addition to
the countless helpful discussions.

\newpage

\appendix{\bf APPENDIX A Example of a large-$n$ asymptotic calculation}

In this appendix, we will show in some detail how to obtain
the asymptotic expressions of the various integrals in
the exponent, in the non-perturbative regime, $n>{Q^2 \over \Lambda^2}$.
As an example
we will work out the integral $J_1$, eq.~(\ref{fivethreea}).
Performing the $y$-integration, we have
\begin{equation}
J_1(n,t)={b_3\over b_2^2}\biggl\{J_{1;1}(n,t/2)-J_{1;1}(n,t)
+J_{1;2}(n,t/2)-J_{1;2}(n,t)\biggr\},
\label{appoo}
\end{equation}
where, after a change of variables $\zeta\to 1-\zeta$,
\begin{equation}
J_{1;1}(n,t)=\int_P {d\zeta\over \zeta}W(n,\zeta){\ln(1+(1/t)\ln\zeta)\over
1+(1/t)\ln\zeta},
\label{appot}
\end{equation}
and
\begin{equation}
J_{1;2}(n,t)=\int_P{d\zeta\over \zeta}W(n,\zeta){1\over 1+(1/t)\ln\zeta}.
\label{appoth}
\end{equation}
$W(n,\zeta)$ is given by eq.~(\ref{weight}).

Focusing on the first of these functions we can write, after the discussion
in sec. 4, its large-$n$ approximation as
\begin{equation}
J_{1;1}(n,t)_>\simeq -\int_{{\cal P}_0}{d\zeta\over \zeta}{\ln(1+(1/t)
\ln\zeta)\over 1+(1/t)\ln\zeta},
\label{appof}
\end{equation}
where the contour ${\cal P}_0$ extends from $n_0(n)$ to $1$ above
and below the real axis and comprises a straight section and
a semicircle around the singularity, with radius $\delta$, see fig. 1a.

The semicircle contributions to the integral are
\begin{eqnarray}
{1\over 2}J_{1;1}(\cap)+{1\over 2}J_{1;1}(\cup)=
&-{\rm Re}\left\{\int_\pi^0{i\delta{\rm e}^{i\theta}d\theta\over \zeta_t+
\delta{\rm e}^{i\theta}}{\ln(1+(1/t)\ln\zeta_t+(\delta/t\zeta_t){\rm e}^
{i\theta})\over 1+(1/t)\ln\zeta_t+(\delta/t\zeta_t)
{\rm e}^{i\theta}}\right\}\nonumber \\
=&-{\rm Re}\left\{it\int_\pi^0 d\theta\left[\ln\left({\delta\over t\zeta_t}
\right)+i\theta\right]\right\}=
-{t\pi^2\over 2}\ .
\label{appofive}
\end{eqnarray}
On the other hand, the straight-line contributions are
\begin{eqnarray}
{1\over 2}J_{1;1}(+i\epsilon)+{1\over 2}J_{1;1}(-i\epsilon)
=&-{\rm Re}\left\{\int_{n_0(n)}^{\zeta_t-\delta}{dx\over x}{\ln(|1+(1/t)\ln x|
{\rm e}^{i\pi})\over 1+(1/t)\ln x}\right\}-
\int_{\zeta_t+\delta}^1{dx\over x}{\ln(1+(1/t)\ln x)\over
1+(1/t)\ln x}\nonumber \\
=&-{t\over 2}\ln^2\left({\delta\over t\zeta_t}\right)+{t\over 2}\ln^2\left[
{1\over t}\ln\left({1\over n_0(n)}\right)-1\right]+{t\over 2}\ln^2\left(
{\delta\over t\zeta_t}\right)\ .
\label{apposix}
\end{eqnarray}
 From eqs.~(\ref{appofive}), (\ref{apposix}), we obtain the desired asymptotic
formula for the integral in the non-perturbative regime:
\begin{equation}
J_{1;1}(n,\alpha)_>\simeq {t\over 2}\left\{\ln^2\left[{1\over t}\ln\left(
{1\over n_0(n)}\right)-1\right]-\pi^2\right\}\equiv F(n,t)\ .
\label{apposeven}
\end{equation}
Notice that the functional form of the divergent $n$-behavior in
eq.~(\ref{apposeven}), in terms of the quantity
$$L(n,t)\equiv {1\over t}\ln\left({1\over n_0(n)}\right)-1,\\ $$
which contains the perturbative/nonperturbative interface, is
dictated by the structure of the IR singularities in eq.~(\ref{apposix})\
(which, of course, cancel).
The rest of the integrals in sec. 4 can be worked out in exactly the same
way.

\newpage

\appendix{\bf FIGURE CAPTIONS}

\begin{description}
\item{Figure 1.} Contours for evaluating the Principal-Value exponent.
\begin{description}
\item{(a)} General definition.
\item{(b)} Principal-Value contour proper, ${\cal P}$.
\item{(c)} ``Box" contour $\bar{P}$, for numerical evaluation.
\end{description}
\item{Figure 2.} Exact and perturbative exponents.
\begin{description}
\item{(a)} $Q=5{\rm GeV}$: Dot=$E(n,\alpha)_L$; Dash=$E(n,\alpha,{\cal
N})_L$;\\  Solid=$E(n,\alpha)$; Dot-dash=$E(n,\alpha,{\cal N})$.
\item{(b)} Same as (a) but for $Q=10{\rm GeV}$.
\item{(c)} Same as (a) but for $Q=90{\rm GeV}$.
\end{description}
\item{Figure 3.} Exact and nonperturbative (large-$n$) exponents.
\begin{description}
\item{(a)} $Q=5{\rm GeV}$: Dot=$E(n,\alpha)_L$; Dash=$E(n,\alpha)_{L>}$;\\
Solid=$E(n,\alpha)$; Dot-dash=$E(n,\alpha)_>$.
\item{(b)} Same as (a) but for $Q=10{\rm GeV}$.
\item{(c)} Same as (a) but for $Q=90{\rm GeV}$.
\end{description}
\item{Figure 4a.} Exact and perturbative next-to-leading exponents.\\
Solid=$E(n,0.075)_{NL}$; Dot=$E(n,0.075,{\cal N})_{NL}$;
Short dash=$E(n,0.061)_{NL}$; \\ Long dash=$E(n,0.061,{\cal N})_{NL}$;
Dot-short dash=$E(n,0.039)_{NL}$;\\ Dot-long dash=$E(n,0.039,{\cal N})_{NL}$.
\item{Figure 4b.} Exact and non-perturbative (large-$n$) next-to-leading
exponents.\\
Notation as in 4a, but with the substitution
$E(n,\alpha,{\cal N})_{NL}\to E(n,\alpha)_{NL>}$.
\item{Figure 5.} The resummed hard function ${\cal H}(z,\alpha)$.
\begin{description}
\item{(a)} $Q=5{\rm GeV}$: Dot=${\cal H}(z,\alpha)_L$;
Dash=${\cal H}(z,\alpha,{\cal N})_L$; \\
Solid=${\cal H}(z,\alpha)$; Dot-dash=${\cal H}(z,\alpha,{\cal N})$.
\item{(b)} Same as (a) but for $Q=10{\rm GeV}$.
\item{(c)} Same as (a) but for $Q=90{\rm GeV}$.
\end{description}
\end{description}

\end{document}